\documentclass[preprintnumbers,preprint,prd,floatfix]{revtex4}
\usepackage{amssymb,amsmath}
\usepackage{graphicx,color}

\newcommand{\bra}[1]{\big<#1\big|}
\newcommand{\ket}[1]{\big|#1\big>}
\newcommand{\gm}[1]{\gamma^{#1}}
\newcommand{\rd}{\partial}

\begin{document}
\title{$I=2$ $\pi$-$\pi$ scattering length
		with dynamical overlap fermion }

\author{Takuya Yagi$^{\;a,b}$,
        Shoji Hashimoto$^{\;a,c}$,
        Osamu Morimatsu$^{\;a,b,c}$,
        Munehisa Ohtani$^{\;a,d}$}

\affiliation{
{$^{a}$}Institute of Particle and Nuclear Studies, High Energy Accelerator Research Organization
(KEK), Tsukuba 305-0801, Japan\\
\llap{$^{b}$}Department of Physics, University of Tokyo, Tokyo 113-0033, Japan\\
\llap{$^{c}$} School of High Energy Accelerator Science,
the Graduate University for Advanced Studies 
 (Sokendai), Tsukuba 305-0801, Japan\\
\llap{$^{d}$} %
Physics Department, School of Medicine, Kyorin University, Tokyo 181-8611, Japan
}

\begin{abstract}
We report on a lattice QCD calculation of the $I=2$ $\pi\pi$
scattering length using the overlap fermion formulation for both sea
and valence quarks.
We investigate the consistency of the lattice data with
the prediction of the next-to-next-to-leading order chiral perturbation theory
after correcting finite volume effects.
The calculation is performed on gauge ensembles of two-flavor QCD
generated by the JLQCD collaboration on a $16^3\times 32$ lattice at a
lattice spacing $\sim$ 0.12~fm. 
\end{abstract}

\preprint{KEK-CP-257}
\preprint{KEK-TH-1489}

\maketitle

\section{Introduction}
In Quantum Chromodynamics (QCD), the pattern of chiral symmetry
breaking in its vacuum governs the interaction among pions, as
dictated by the chiral effective theory \cite{Weinberg:1978kz}.
Beyond the limit of massless soft pions, the pion interaction receives
corrections due to finite pion mass and momentum, which can be
systematically calculated using the framework of chiral perturbation
theory (ChPT) \cite{Gasser:1983yg}.
For the $\pi\pi$ scattering, the two-loop calculation has been
performed and the analysis of experimental data has been attempted so
far \cite{Colangelo:2001df}. 

Although the phenomenological analysis has provided evidence that the
pion scatterings are fitted well by ChPT, it has its own limitation
that the range of applicability of the chiral expansion is not known a
priori. 
Since the quark mass in the nature is a constant, it is simply not
possible to study the mass range that ChPT can be used.
This is relevant to the question whether kaon can also be treated
in ChPT on the equal footing.
Furthermore, so-called the low-energy constants, the parameters
appearing in ChPT, cannot be determined within ChPT.
Eventually, one has to solve QCD in order to make a parameter-free
comparison with the experimental data.

Lattice QCD calculation provides a powerful tool to approach such a
goal. 
As far as the convergence of the chiral expansion is concerned, the
best studied quantities are the pion mass $m_\pi$ and decay constant $f_\pi$
(for a recent review, see for instance \cite{Scholz:2009yz}).
Among other lattice studies, the JLQCD and TWQCD collaborations
investigated the convergence in two-flavor QCD, using the
next-to-next-to-leading order (NNLO) of ChPT formula with various expansion
parameters that differ only beyond next-to-leading orders (NLO)
\cite{Noaki:2008iy}. 
The advantage of this particular work is in the use of the overlap
fermion formulation on the lattice.
Since the overlap fermion preserves exact chiral symmetry (in
massless QCD) while maintaining the flavor symmetry as in the
continuum theory, the application of the chiral effective theory is
justified, in contrast to other lattice studies using the
Wilson-type or staggered lattice fermions, for which some modification
of ChPT with extra unknown constants is mandatory.
The study \cite{Noaki:2008iy} showed that the chiral expansion
converges well up to the kaon mass region as long as the expansion
parameter $\xi\equiv m_\pi^2/(4\pi f_\pi)^2$ is used
where finite quark mass correction is included in $m_\pi$ and $f_\pi$.
With this expansion parameter, one could effectively resum a part of
higher order terms.

Extension of such study to other physical quantities is important in
order to obtain a better idea on the overall applicability of ChPT.
The JLQCD and TWQCD collaborations have so far calculated and analyzed
the electromagnetic and scalar form factors of pion \cite{Aoki:2009qn}
using the NNLO formula in ChPT.
For these quantities to be consistent with the corresponding
phenomenological analysis, it turned out that the NNLO terms have to be
included. 
It is crucial to extend the test of ChPT beyond the simplest
quantities studied so far. 

In this work, we study the $I=2$ $\pi\pi$ scattering length.
The lattice calculation of this quantity is easiest among other hadron
scattering amplitudes, since the quark-flow diagrams are limited to
those without pair-creation or annihilation between initial and final
two-pion states.
One uses the formula provided by L\"uscher to relate the two-pion
state energy with the scattering length, or in general scattering
phase shift \cite{Luscher:1986pf,Luscher:1990ux}.
There have been a number of lattice calculations with and without
quenched approximation
\cite{Sharpe:1992pp,Kuramashi:1993ka,Ishizuka:2004} in the past, 
and recently more realistic calculations with light dynamical quarks
have also been performed \cite{Beane:2007xs,Feng:2009ij}. 
We carry out a similar calculation but using the overlap fermion
formulation on the lattice, that has exact chiral symmetry at finite
lattice spacings.

Our calculation is carried out on a $16^3\times 32$ lattice 
with lattice spacing $a\simeq$ 0.12~fm.
This two-flavor QCD ensemble is generated by the JLQCD and TWQCD
collaborations \cite{Aoki:2008tq} using the overlap fermion
formulation for sea quarks.
The sea quark mass ranges from $m_{\rm s}/6$ to $m_{\rm s}$ 
with $m_{\rm s}$ the physical strange quark mass.
In order to maintain exact chiral symmetry, the global topological
charge $Q$ of the SU(3) gauge field is fixed to its initial value in
the simulation, which is typically zero.
This is irrelevant in the infinite volume limit as the global
topological charge should not affect local physics, but induces finite
volume effect that scales as $1/V$ at finite space-time volume $V$
\cite{Aoki:2007ka}.

Since the lattice volume $(1.9\mbox{~fm})^3$ is not as large as those
used in previous  studies, {\it e.g.} \cite{Beane:2007xs,Feng:2009ij},
finite volume effects have to be carefully investigated.
The effect due to the fixed global topology can be corrected following
the strategy outlined in \cite{Aoki:2007ka}.
The conventional finite volume effect due to pions wrapping around
the lattice can also be estimated using ChPT, as calculated for pion mass 
and decay constant \cite{Colangelo:2005gd,Colangelo:2006mp}.
In this work we use the formula developed in \cite{Bedaque:2006yi} for
the $\pi\pi$ scattering.

This paper is organized as follows. 
After describing the details of the simulation setup in the next
section, we discuss the method to extract the two-pion state energy in
our calculation in Section~\ref{sec:Correlation_function}.
In Section~\ref{sec:Scattering_length} we show numerical results of our
simulations and discuss the correction for finite volume effects.

\section{Lattice setup}
We use the overlap fermion formulation 
\cite{Neuberger:1997fp,Neuberger:1998wv}
for both sea and valence quarks.
Its Dirac operator for a massive quark is written as
\begin{equation}
  \label{eq:ov_operator}
  D(m) = \left(m_0+\frac{m}{2}\right) + \left(m_0-\frac{m}{2}\right)
  \gamma_5 \mathrm{sgn}[H_W(-m_0)], 
\end{equation}
where $H_W(-m_0)$ is the standard hermitian Wilson-Dirac operator with
a large negative mass term.
We choose $-m_{0}=-1.6$ in the lattice unit.
The physical quark mass is controlled by $m$.
For the gauge sector, we use the Iwasaki gauge action \cite{Iwasaki:1985}
at $\beta = 2.30$ together with extra (irrelevant) Wilson fermions
that suppress the near-zero modes of $H_W(-m_0)$ \cite{Fukaya:2006vs},
so that the fermionic determinant does not have singularity and the
Hybrid Monte Carlo (HMC) simulation becomes feasible.

The numerical simulation is performed on a $16^3\times 32$ lattice 
with two flavors of dynamical fermions
as one of the main projects of the JLQCD and TWQCD collaborations
\cite{Aoki:2008tq}.
The lattice spacing $a$ is determined as 0.1184(3)(17)(12)~fm 
from the Sommer scale $r_{0}=$ 0.49~fm \cite{Sommer:1993ce}.
The sea and valence quark masses are set to 
0.015, 0.025, 0.035, 0.050, 0.070, and 0.100 in the lattice unit,
which correspond to the mass range between $m_{\rm s}/6$ and $m_{\rm s}$
with $m_s$ the physical strange quark mass.

For each sea quark mass, we calculate the two-pion state energy on 
about 200 gauge configurations, each separated by 50 HMC trajectories.
(The number of the configurations used in the analysis is
191, 193, 187, 193, 193 and 187 for the sea quark mass
0.015, 0.025, 0.035, 0.050, 0.070 and 0.100, respectively.)
The JLQCD collaboration has calculated and stored the lowest 50 pairs of
eigenvalues and eigenvectors of the overlap-Dirac operator for their
dynamical configurations.
We utilize them to precondition the overlap solver, which makes the
calculation faster by an order of magnitude \cite{Aoki:2008tq}. 

On these ensembles, the global topological charge $Q$ is fixed to
zero during the HMC simulations in order to suppress the occurrence of
the unphysical near-zero modes of $H_W(-m_0)$.
This makes the application of the overlap operator $D(m)$ much faster.
The effect of artificially fixed topological charge on the physical
quantities can be understood and indeed be corrected as a finite
volume effect of $O(1/V)$ \cite{Aoki:2007ka}, which will be discussed
later.

\section{Correlation functions}
\label{sec:Correlation_function}
In order to extract the $\pi\pi$ scattering length through
L\"uscher's formula \cite{Luscher:1986pf,Luscher:1990ux}, 
we need the energies of $\pi$ and $\pi\pi$ system in a finite volume. 
The energy of a hadron state $h$ is extracted from the correlation
function $C_{h}(t;t_{0})$,
which describes the temporal propagation of the state from $t_{0}$ to $t$. 
The source and sink operators are chosen such that they have some
overlap with the state of interest.
In our calculation, we consider
$C_{\pi}(t;t_{0})$ and $C_{\pi\pi}(t;t_{0})$ defined by
\begin{eqnarray}
  \label{eq:corr2}
  C_{\pi}(t;t_{0}) &=& \frac{1}{L^3}\sum_{\vec{x}}
  \bra{0}{\pi(t,\vec{x})}^{\dagger}W(t_{0})\ket{0},
  \\
  \label{eq:corr4}
  C_{\pi\pi}(t;t_{0}) 
  &=& \frac{1}{L^3}\sum_{\vec{x},\vec{y}}	
  \bra{0}
  \left(
    \pi(t,\vec{x})\pi(t,\vec{y})
  \right)^{\dagger}
  W(t_{0})W(t_{0})\ket{0},
\end{eqnarray}
where $\pi(t,\vec{x})$ is an interpolating field of the pion, a local
pseudo-scalar density.
The source operator $W(t_0)$ corresponds to a wall source spread over
the space. 
For the calculation of the scattering length, only the zero-momentum
states are necessary; the correlators eq.~(\ref{eq:corr2}) and
eq.~(\ref{eq:corr4}) are projected onto the zero-momentum states.

The operators $\pi(t,\vec{x})$ and $W(t)$ are written in terms of
quark fields $u$ and $d$ as 
\begin{eqnarray}
  \label{eq:inter_polating}
  \pi(t,\vec{x})
  &=& \bar{d}(t,\vec{x})\gm{5}u(t,\vec{x}), \\
  W(t) 		
  &=& \sum_{\vec{x},\vec{y}}\bar{d}(t,\vec{x})\gm{5}u(t,\vec{y}).
\end{eqnarray}
The wall source is used on gauge configurations fixed to the Coulomb
gauge.

\begin{figure}
  \begin{center}
    \includegraphics[width=8cm]{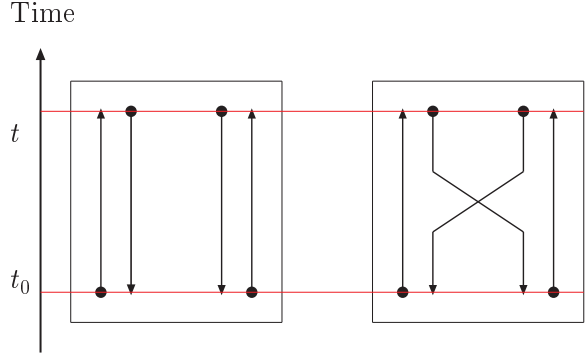}
    \caption{
      The Wick-contracted quark-line diagrams for the four-point function of
      the $I=2$ $\pi\pi$ scattering.
      Only the diagrams propagating forward in time are shown for simplicity.
      Each line represents a quark propagator $<q(x) \bar{q}(y)>$ where $q$
      is either $u$ quark or $d$ quark.
      One end of the line with a dot represents a source $\bar{q}(y)$ and
      the other end of the line with an arrow represents a sink $q(x)$.
      The left and right diagrams are called ``direct" and ``crossed", respectively.
    }
    \label{fig:topology}
  \end{center}
\end{figure}

For the $I=2$ scattering amplitude we can consider a scattering of two
$\pi^+$'s, so that no creation or annihilation of quarks occur between
$t_0$ and $t$.
The Wick-contracted quark-line diagrams are shown in
Figure~\ref{fig:topology}.
In addition to the ``direct'' contribution where the two-pion states
do not exchange valence quarks (left), there is a valence-quark
exchange diagram, which is called ``crossed'' (right).

Since the temporal extent of the lattice is finite, the correlation
functions are not simply an exponential function,
$C(t;0)\sim\exp(-m_\pi t)$,
decaying in time by a rate $m_\pi$,
but have a structure that reflects the boundary condition.
For the periodic boundary condition in the temporal direction, the
two-point function has the form 
$\exp(-m_\pi t)+\exp(-m_\pi(T-t))$,
where the second term describes the state propagating backward in
time. 
Here, $T$ is the temporal extent of the lattice.

\begin{figure}
  \begin{center}
    \includegraphics[width=16cm]{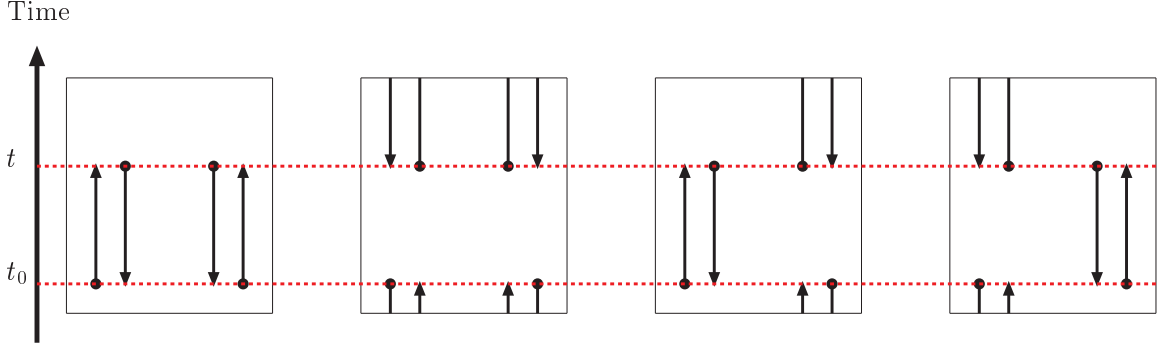}
    \caption{
      Four ``direct'' Wick-contracted quark-line diagrams for the four-point function of
      the $I=2$ $\pi\pi$ scattering, which exist due to the periodic boundary condition
      in the temporal direction.
    }
    \label{fig:wrap-around}
  \end{center}
\end{figure}

For the case of the four-point function, each pion may propagate in
forward or backward directions in time, so that there are four
distinct contributions to the correlation function as shown in
Figure~\ref{fig:wrap-around}.
Two of them contain the two-pion state, while the others contain only one
pion at a given time-slice.
Thus, the four-point function $C_{\pi\pi}(t;0)$ may have a form
\begin{eqnarray}
  \label{eq:4pt}
  C_{\pi\pi}(t;t_{0})
  &\sim& 
  \alpha\left(e^{-E_{\pi\pi}t}+ e^{-E_{\pi\pi}(T-t)}\right) 
  +\beta e^{-m_{\pi}t}e^{-m_{\pi}(T-t)} 
  \nonumber\\
  &=& 2\alpha\cosh\left(-E_{\pi\pi}(t-T/2)\right) + \beta e^{-m_\pi T},
  \label{eq:4pt-II}
\end{eqnarray}
where $\alpha$ and $\beta$ are constants.
$E_{\pi\pi}$ represents the two-pion state energy.
The second term in eq.~(\ref{eq:4pt}) represents the effect from two pions
propagating in opposite directions in time
and is called the wrap-around term since one pion wraps around the temporal direction.
It amounts to a constant contribution.
Although this effect is suppressed for large $T$ as $\exp(-m_\pi T)$,
its contamination could become non-negligible in the middle of the
lattice, since the two-pion state signal is also suppressed as
$\exp(-E_{\pi\pi}T/2)\simeq\exp(-m_\pi T)$.
We therefore fit the two-pion correlator by eq.~(\ref{eq:4pt}).

In order to identify the time-separations where the ground state
dominates, we consider the following ratios:
\begin{eqnarray}
  \label{eq:R_p}
  R_{\pi}(t) &=& \frac{C_{\pi}(t+1)}{C_{\pi}(t)}
  \nonumber\\
  & \to & 
  \frac{\cosh\left(-m_{\pi}(t+1-T/2)\right)}{
    \cosh\left(-m_{\pi}(t-T/2)\right)},
  \\
  \label{eq:R_pp}
  R_{\pi\pi}(t) &=& 
  \frac{C_{\pi\pi}(t+1)-C_{\pi\pi}(t)}{C_{\pi\pi}(t)-C_{\pi\pi}(t-1)}
  \nonumber\\
  & \to & 
  \frac{
    \cosh\left(-E_{\pi\pi}(t+1-T/2)\right) - 
    \cosh\left(-E_{\pi\pi}(t-T/2)\right)}{
    \cosh\left(-E_{\pi\pi}(t-T/2)\right) - 
    \cosh\left(-E_{\pi\pi}(t-1-T/2)\right) 
  }.
\end{eqnarray}
At large time separations, these ratios approach the form given in the
second line of each equation, as the ground state dominates.
From these ratios, we may extract the effective mass $m_\pi(t)$ and
effective two-pion energy $E_{\pi\pi}(t)$ for each time slice.
By inspecting the resulting effective energies, we are able to
identify the region where the ground state dominates.

In the actual simulation, we adopt a technique called low mode average (LMA),  \cite{Giusti:2004yp,DeGrand:2004qw} for the correlation function.
In this technique the quark propagator is separated into the eigenmodes of the overlap-Dirac operator. 
Then, a part of the correlation function, in which only low modes of the quark propagator participate, is averaged over the time of the source, as shown in Appendix~\ref{app:low_mode_average}.
By taking an average we expect to gain statistics and have a more stable plateau in the correlation function of pions.

Showing numerical results of the effective energy,
here we discuss the impacts of the LMA and the wrap-around effects on the correlation functions.

For the two-point function, it is found that the correlation function with LMA has smaller statistical errors than the one without LMA for all quark mass.
This is consistent with previous results \cite{Noaki:2007es}.
For the four-point function,
we show in Figure~3 the two-pion effective energy obtained from eq.~(\ref{eq:R_pp}) with and without LMA.
The results show that LMA works well for small quark masses, $m = 0.015 - 0.050$, 
but it is not so effective for larger quark masses.
Especially, for $m = 0.100$, the data show larger statistical errors with LMA.
We therefore use the effective energy without LMA at the heaviest quark mass
for the following analysis.

\begin{figure}
  \begin{center}
	\scalebox{0.95}{
   \includegraphics{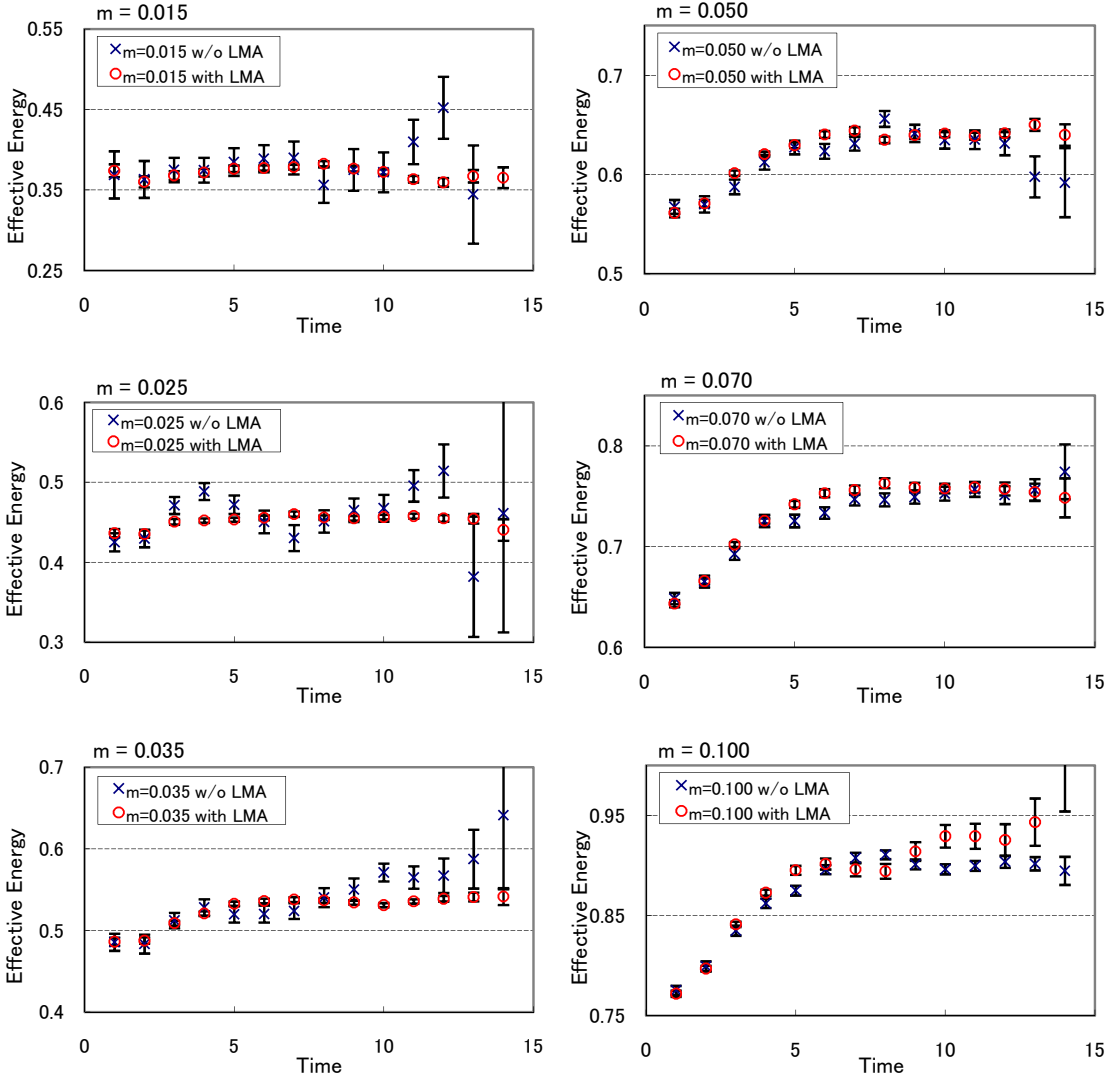}}
    \caption{
Comparison of the two-pion effective energy $E_{\pi\pi}(t)$
with LMA (circles) and without LMA (crosses) at each quark mass.
For all quark masses $E_{\pi\pi}(t)$ is extracted with the wrap-around term.
    }
    \label{fig:2}
  \end{center}
\end{figure}
\begin{figure}
  \begin{center}
	\scalebox{0.95}{
   \includegraphics{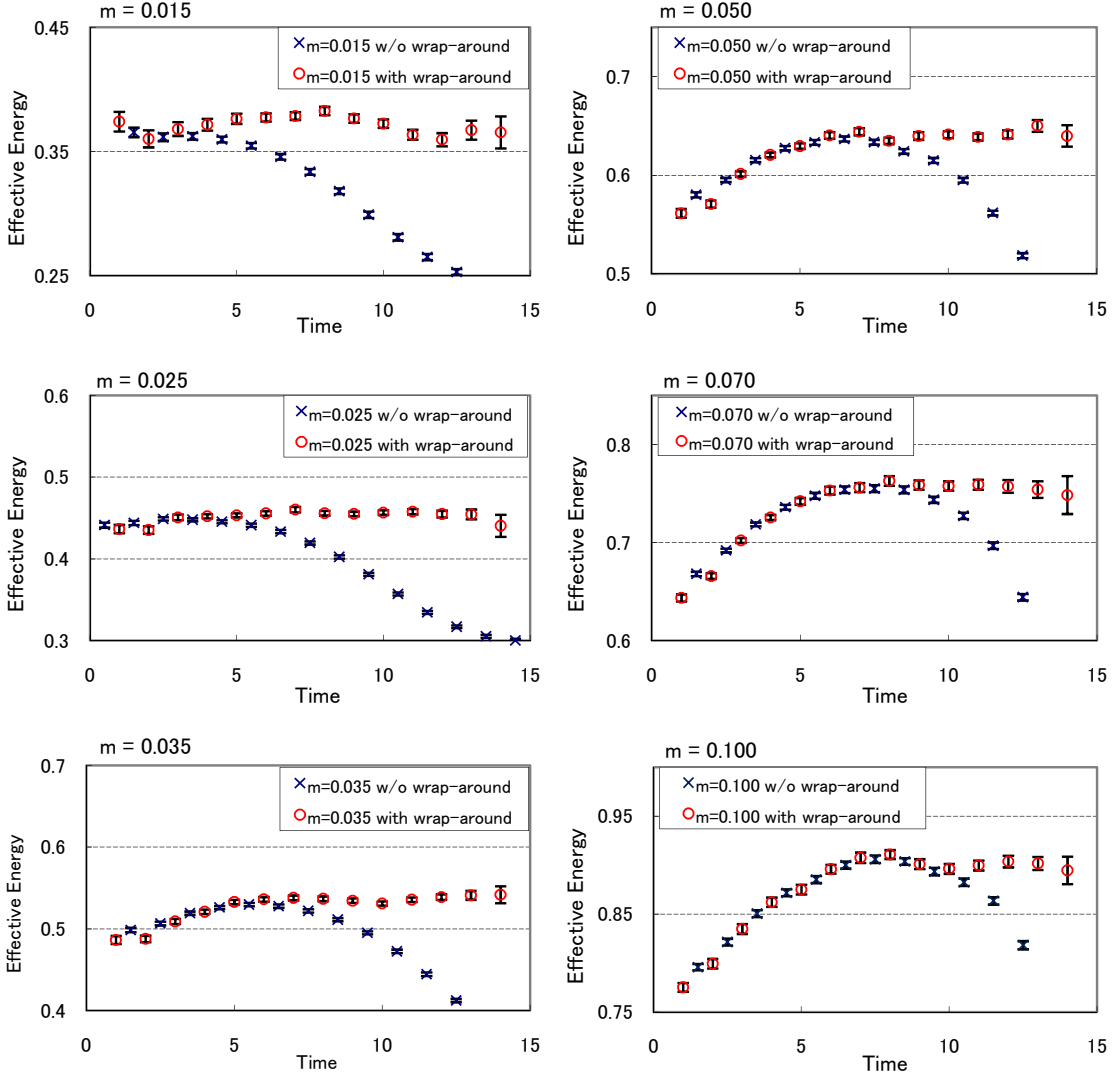}}
    \caption{
Comparison of the effective two-pion energy $E_{\pi\pi}(t)$
with the wrap-around term (circles) and without the wrap-around term (crosses)
at each quark mass.
For $ma = 0.100$ $E_{\pi\pi}(t)$ is extracted without LMA while for other quark masses
 $E_{\pi\pi}(t)$ is extracted without LMA.
}
    \label{fig:3}
  \end{center}
\end{figure}

In Figure~4, we compare the effective energy extracted from the four-point function
with and without taking account of the wrap-around term.
The latter is done by fitting the effective energy only with the hyperbolic cosine term, the first term of eq.~(\ref{eq:4pt-II}).
On the one hand the effective energy with the wrap-around term shows plateaus,
while the effective energy without the wrap-around term rapidly decreases as the time increases.
On the other hand the effective energy with the wrap-around term has larger statistical errors near
$t \sim T/2$.
This can be understood as follows.
As $t$ approaches $T/2$, the hyperbolic cosine behaves like a constant.
Due to this nearly degenerate behavior, the errors become large when one fits the correlation functions near $t \sim T/2$ to the sum of the hyperbolic cosine and the constant terms.

As the quark mass becomes smaller, the effective energy without the wrap-around term starts to 
decrease earlier in time.
It is, therefore, absolutely necessary to use the effective energy with the wrap-around term for 
smaller quark masses, $m = 0.015 - 0.050$.
For larger quark masses, $m = 0.070$ and $0.100$, however, the situation is different. 
The time region near $t \sim T/2$ turns out unimportant due to large statistical errors when one 
extracts the ground state energy, $E_{\pi\pi}$, from the effective energy with the wrap-around term.
In the time region, which is important in determining $E_{\pi\pi}$, the effect of the wrap-around term is small.
For $m = 0.070$ and $0.100$, it turns out that $E_{\pi\pi}$ determined from the effective energy without the wrap-around term in the range $6 \le t \le 9$ coincide with the one with the wrap-around term.
In later analysis we use the former.

In summary, using LMA for the whole quark masses except $m = 0.100$ 
and including the wrap-around terms except $m = 0.700$ and $0.100$,
we fit the correlation functions in the time range $9 \le t \le 15$ and obtain 
the ground state energies $m_{\pi}$ and $E_{\pi\pi}$.
For $m = 0.100$, the effective energy is extracted without the LMA,
and the time range is chosen to be $6 \le t \le 9$ for $m = 0.070$ and $0.100$ 
so that the effective energy coincides with and without the wrap-around term.
The results are shown in Table~\ref{tab:results_1} together with the statistics, i.e.\ the number of 
configurations used in the calculation of two- and four-point functions.  
\begin{table}
  \begin{center}
    \caption{Summary of the quark mass, $m_{q}$, the ground state energies, $m_{\pi}(L)$ and $E_{\pi\pi}$ extracted from and two- and four-point functions and the statistics.
    We note $m_{\pi}(L)$ in order to distinguish it from the pion mass including the finite volume effect, $m_\pi$.}
    \begin{tabular}{|c|c|c|c|}
	\hline
$m_{q}$& $m_{\pi}(L) $ & $E_{\pi\pi}$ & statistics\\
	\hline  
	0.015& 0.1716(12)&0.3697(38)&191\\
	0.025& 0.21984(87)&0.4565(22)&193\\
	0.035& 0.26104(84)&0.5345(19)&187\\
	0.050& 0.31332(80)&0.6407(25)&193\\
	0.070& 0.37101(85)&0.7547(32)&193\\
	0.100& 0.4475(15)&0.9033(31) &187\\
	\hline
    \end{tabular}
	\label{tab:results_1}
  \end{center}
\end{table}

\section{Scattering length}
\label{sec:Scattering_length}
Next, we explain how we extract the scattering length at the physical pion mass from the energies $m_\pi$ and $E_{\pi\pi}$ in a finite volume obtained in the previous section.

\subsection{L\"uscher's Formula}
L\"{u}scher's formula \cite{Luscher:1986pf,Luscher:1990ux} 
relates the energy of two hadrons in a box of size $L$ with the scattering phase shift of two hadrons $\delta$.
For the two-pion case, the relation is
\begin{equation}
k\,\textrm{cot}\delta = 
		\frac{1}{\pi L}S\left(\frac{kL}{2\pi}\right)
		\label{luc}.
\end{equation}
$k$ is the pion momentum in the center-of-mass system
and is related to the $\pi\pi$ energy, $E_{\pi\pi}$, as $E_{\pi\pi}= 2\sqrt{k^2 + m^2_{\pi}}$. 
$E_{\pi\pi}$ is measured on the lattice in the center-of-mass system, obtained in the previous section.
The function $S$ is defined as follows \cite{Beane:2003yx}:
\begin{align}
S\left(\frac{k L}{2\pi }\right) 
	&= 4 \pi^2 L 
	\left[
	\sum_{\vec{q}=2\pi\vec{n}/L}
	- \int\frac{d^3 q}{(2\pi)^3}
	\right]
	  \frac{1}{\vec{q}^2 - \vec{k}^2} \\
	&=\lim_{\varLambda\rightarrow \infty}
	\sum_{|\vec{n}| < \varLambda }
	\frac{1}{\vec{n}^2 - \frac{k^2 L^2}{4\pi^2}}
	- 4\pi\varLambda.
\end{align}
Near the $\pi\pi$ threshold, the inverse functions of $S$ can be expanded in terms of $1/L$ 
and it leads to the following expression
\begin{align}
  \label{eq:luscher}
E_{\pi\pi} - 2m_{\pi}  = -\frac{4\pi a_{\pi\pi}}{m_{\pi}L^{3}}
        \left\{
        1 + c_{1}\frac{a_{\pi\pi}}{L} 
	+ c_{2}\left(\frac{{a_{\pi\pi}}}{L}\right)^{2}
        \right\}
        + {\cal O}(L^{-6}),
\end{align}
where $a_{\pi\pi}$ is the $\pi\pi$ scattering length defined by
\begin{align}
 \frac{1}{a_{\pi\pi}} = \lim_{k \rightarrow 0} k\cot\delta.
\label{eq:a-delta}
\end{align}
The numerical constants $c_1$ and $c_2$ are 
$-2.837297$ and $6.375183$, respectively \cite{Luscher:1986pf}. 

\subsection{Finite Volume Effects}
Through L\"uscher's formula eq.~(\ref{eq:luscher}),
we can extract the scattering length $a_{\pi\pi}$ from $m_{\pi}$ and $E_{\pi\pi}$ measured on the lattice.
However, there are corrections neglected in deriving eq.~(\ref{eq:luscher}), i.e.\ the finite volume effects.
These corrections are small if the volume is large enough.
But the spatial size of our lattice $L=16 \times 0.1184 \ \mbox{fm} \sim 1.9 \ \mbox{fm}$, which is not large enough to neglect these corrections a priori.
We therefore investigate the impact of finite volume effects in our data.
We investigate the conventional effect due to pions wrapping around the lattice and also the effect due to the fixed global topology.

\subsubsection{Analytical Formula}
Since the periodic boundary condition is imposed in the spacial direction of the lattice,
particles can wrap around the lattice and quantum fluctuations of the virtual particles, i.e.\
loop contributions, to the physical quantities are modified.
The modifications induce corrections of $O(\exp(-mL))$ to physical quantities
measured in the finite box of size $L$, 
where $m$ is the mass of the virtual particle.
The corrections are dominated by the lightest particle, the pion, 
and become more important as the pion mass becomes smaller in the simulation. 
These pion-loop corrections can be estimated using ChPT.

We need to know the correction for the $\pi\pi$ scattering length
$\Delta a_{\pi\pi}$ as well as the correction for the pion mass $\Delta m_\pi$.
$\Delta m_\pi$ is known at NNLO of ChPT \cite{Noaki:2008iy}
while $\Delta a_{\pi\pi}$ is obtained as described below.

\begin{figure}
  \begin{center}
  \scalebox{0.60}{
   \includegraphics{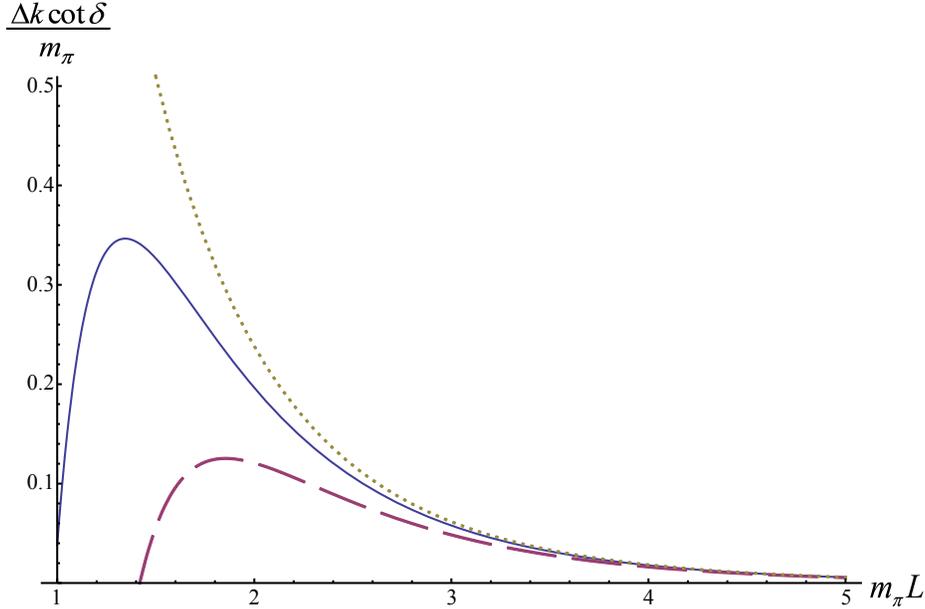}}
    \caption{$\Delta (k \cot \delta)/m_\pi$ vs. $m_\pi L$. 
	The solid line represents the exact formula eq.~(\ref{eq:correction_1}) 
	while the dashed and dotted lines represent the expanded formulae with first two and three \
	terms in eq.~(\ref{eq:correction_2}), respectively.
	}
    \label{fig:1}
  \end{center}
\end{figure}

From the $\pi\pi$ scattering length in a finite box $a_{\pi\pi}(L)$
the $\pi\pi$ scattering length in the infinite volume $a_{\pi\pi}$ 
can be obtained as
\begin{align}
m_{\pi}a_{\pi\pi}=
 \frac{m_{\pi}(L)a_{\pi\pi}(L)}{1 +m_{\pi}(L) a_{\pi\pi}(L)
\cdot
\displaystyle{\lim_{k\rightarrow 0}\frac{\Delta (k\cot\delta)}{m_{\pi}}}},
	\label{eq:distorted1}
\end{align}
where $\Delta (k\cot\delta)$ is the correction for $k\cot\delta$ in a finite 
box, 
\begin{align}
\Delta (k\cot\delta)(L) = k\cot\delta - k\cot\delta(L).
	\label{eq:distorted2}
\end{align}
The formula for $\Delta (k\cot\delta)$
is calculated at the NLO 
of ChPT in ref.~\cite{Bedaque:2006yi},
which is expanded in $(m_{\pi}L)^{-1}$ in the low-momentum limit as
\begin{align}
		\lim_{k \rightarrow 0} \frac{\Delta(k\cot\delta)(L)}{m_{\pi}}
		&=8\pi
	\left[
		\frac{\rd}{\rd m_{\pi}^2} i\Delta \mathcal{I}(m_{\pi})
		+2i \Delta \mathcal{J}_{\rm exp}(4m_{\pi}^2) 
	\right] 
	\label{eq:correction_1}\\
		&= -\frac{1}{\sqrt{2\pi}}
		\sum_{|\vec{n}| \neq 0}
		c(n)\frac{e^{-|\vec{n}| m_{\pi} L}}
		{\sqrt{|\vec{n}| m_{\pi} L }}
		\left[
		1 - \frac{17}{8}\frac{1}{|\vec{n}| m_{\pi} L} + 
	                  \frac{169}{128}\frac{1}{|\vec{n}|^2 m_{\pi}^2 L^2} + 
		{\cal O}\left(\frac{1}{|\vec{n}|^3 m_{\pi}^3 L^3}\right)
		\right].
	\label{eq:correction_2}
\end{align}
$\Delta\mathcal{I}$ and $\Delta\mathcal{J_{\rm exp}}$ are defined with
the modified Bessel and Struve functions as in ref.~\cite{Bedaque:2006yi}, and
$c(n)$ is the multiplicity factor, i.e.\ how many times
the discretized momentum with $n=|\vec{n}|$ appear in the sum.
In Figure~\ref{fig:1} we show $\Delta (k\cot\delta)(L)/m_{\pi}$ as a function of $m_{\pi} L$.
In this figure we show exact $\Delta (k\cot\delta)(L)/m_{\pi}$, eq.~(\ref{eq:correction_1}),  and also 
$\Delta (k\cot\delta)(L)/m_{\pi}$ expanded in $(m_{\pi}L)^{-1}$ with first two and three terms,
respectively, as in eq.~(\ref{eq:correction_2}).
From this figure, it is seen that the expansion of $\Delta (k\cot\delta)(L)/m_{\pi}$
does not converge well in the region $m_{\pi}L < 3$.
Therefore we use eq.~(\ref{eq:correction_1}) instead of the expanded
form, eq.~(\ref{eq:correction_2}), when we estimate the finite volume effect of virtual pion loops.

In addition to these pion-loop corrections, we also consider
the finite volume effect due to the fixed global topology.
This effect causes corrections of $O(1/V)$ to the Green functions in general
and can be estimated once the topological susceptibility and the $Q$ dependence of
the physical quantity of interests are known \cite{Brower:2003yx,Aoki:2007ka}.
The topological susceptibility has been calculated for the JLQCD gauge configurations
recently.
The $Q$ dependence can be obtained through ChPT.
At the leading order of ChPT, only the pion mass depends on $Q$.
In the present analysis, we therefore include this effect on
the pion mass as done in \cite{Noaki:2008iy}.
Then, the effect on the scattering length is calculated at the NLO of ChPT as
explained in Appendix~\ref{app:global_topology}.
The correction for $E_{\pi\pi} - 2m_{\pi}(L)$ due to the fixed global topology
can be expressed as
\begin{align} 
\delta(E_{\pi\pi} - 2m_{\pi}(L)) =
\frac{1}{4f^2 L^3}
\left[
\frac{1}{8V \chi_{t}} 
 \frac{m_{\pi}^2(0)}{16 \pi^2 f^2 }
\left\{\frac{7}{2}
\left(
 \ln\frac{m_{\pi}^2 (0)}{\mu^2} +1
 \right)
 + l'_{a}(\mu)
\right\}
  + {\cal O}(V^{-2})
\right].
\label{fse_t}
\end{align}
Here $f$ is the pion decay constant in the chiral limit, $\chi_t$ is
the topological susceptibility, $l'_{a}$, which is defined in Appendix~\ref{app:global_topology}, is a combination of
the low-energy constants $\tilde{l}_i$ at the renormalization scale $\mu$,
$V$ is the four-dimensional volume
$V = L^3 \times T$, and $m_{\pi}(0)$ is the pion mass at $\theta=0$.
We obtain $m_{\pi}(\theta=0)$ by combining the pion mass calculated by ourselves
for the fixed topology and the correction estimated
by the JLQCD collaboration \cite{Noaki:2008iy,Chiu:2008}.
For $f$ and $\chi_t$, we use the known results calculated on the same
gauge configurations \cite{Noaki:2008iy,Chiu:2008}.
$l'_{a}(\mu)$ is calculated from
the phenomenological values of the low-energy constants \cite{Colangelo:2001df}.

\subsubsection{Numerical Results}
To estimate the finite volume effect, we first
calculate $E_{\pi\pi}-2m_{\pi}(L)$ from the uncorrected $m_{\pi}(L)$ and $E_{\pi\pi}$ (Table~\ref{tab:results_1} ).
Secondly, we include finite volume effect due to the fixed global topology, $\delta\big(E_{\pi\pi} - 
2m_{\pi}(L)\big)$, eq.~(\ref{fse_t}).
From $E_{\pi\pi}-2m_{\pi}(L)$ with corrections due to the fixed global topology taken into account,
the scattering length $a_{\pi\pi}$ is obtained through
L\"uscher's formula, eq.~(\ref{eq:luscher}).
Finally, we take into account the finite volume effect due to the pion-loop corrections,
$\Delta (k\cot\delta)(L)$ using eq.~(\ref{eq:correction_1}),
and obtain $m_{\pi}a_{\pi\pi}$.

In the above procedure we use the corrected pion mass including finite volume effects
on the r.h.s. of eq.~(\ref{fse_t}), eq.~(\ref{eq:luscher}) and eq.~(\ref{eq:correction_1}).
The corrected pion mass, $m_{\pi}$, is given as
\begin{align}
m_{\pi} &= \frac{m_{\pi}(L)}{(1+R_{m})(1+T_{m})},
\label{eq:fse_m}
\end{align}
where $R_{m}$ and $T_{m}$ are
the corrections from the pion loop and the fixed global topology, respectively,
obtained by the JLQCD collaboration \cite{Noaki:2007}.
We summarize each of the corrections in Table~\ref{tab:fse_1}. 
In this table, all the quantities are given in the lattice unit,
and the first and second errors in $1+T_{m}$ are due to the errors of $\chi_{t}$
and the low-energy constants, respectively.

\begin{table}
\begin{center}
\caption{Summary of finite volume effects on the pion mass. $R_{m}$ and $T_{m}$ are the corrections from the pion loop and the fixed global topology, respectively, obtained by the JLQCD collaboration \cite{Noaki:2007}. $m_{\pi}$ is the corrected pion mass.}
\begin{tabular}{ | c | c c | c |}
\hline 
$m_{q}$
&$1+R_{m}$
&$1+T_{m}$ 
&$m_{\pi}$
\\
\hline
0.015& 1.0226(34) &   0.9761(08)(16)  & 0.1718(14)  \\
0.025& 1.0095(19)  &  0.9858(14)(13) & 0.2209(11) \\
0.035& 1.0048(12)  &  0.9889(21)(13) & 0.2626(10) \\
0.050& 1.00188(51) & 0.99357(91)(89) & 0.31480(91) \\
0.070& 1.00075(23) & 0.99515(41)(81) & 0.37245(92)\\
0.100 & 1.000214(70) & 0.9943(07)(11) & 0.4501(16)\\
\hline
\end{tabular}
	\label{tab:fse_1}
\end{center}
\end{table}

In Table~\ref{tab:fse_2} we summarize finite volume effects on $E_{\pi\pi}-2m_{\pi}(L)$
and the scattering length.
From this table, it is seen that the pion-loop corrections, ${\Delta(k\cot\delta)(L)}/{m_{\pi}}$,
increase as the quark mass becomes lighter, 
while the corrections due to the fixed global topology, $\delta\big(E_{\pi\pi} - 2m_{\pi}(L)\big)$,
become important for heavier quark masses. 
However, it turns out from $m_{\pi}(L)a_{\pi\pi}(L)$ and $m_{\pi}a_{\pi\pi}$ that these finite volume effects are not so large, in fact as large as a few
\%, except for the largest quark mass where the effect is about 10 \%.
In Table~\ref{tab:fse_2} we also show the chiral expansion parameter, $\xi=m_\pi^2/(16\pi^2f_\pi^2)$,  which will be used in the chiral extrapolation in the next subsection.
In our notation, the pion decay constant, $f_{\pi}$, should correspond to the experimental value, 
132 MeV, at the physical pion mass.
This should be kept in mind when one compares our results with others' because
in some references a different convention for the pion decay constant is adopted.

%%%
\begin{table}
\begin{center}
\caption{Summary of finite volume effects on $E_{\pi\pi}- 2m_{\pi}$ and the scattering length. $\delta\big(E_{\pi\pi} - 2m_{\pi}(L)\big)$ and $\frac{\Delta(k\cot\delta)(L)}{m_{\pi}}$ are the corrections from the fixed global topology and the pion loop, respectively. $a_{\pi\pi}(L)$ and $a_{\pi\pi}$ are the uncorrected and corrected $\pi\pi$ scattering length, respectively.
The chiral expansion parameter $\xi$ is also shown.}
\begin{tabular}{ |c|c|c c |c c|c|}
\hline 
$m_{q}$
&$E_{\pi\pi} - 2m_{\pi}(L)$
&$\delta\big(E_{\pi\pi} - 2m_{\pi}(L)\big)$
&$\frac{\Delta(k\cot\delta)(L)}{m_{\pi}}$
& $m_{\pi}(L)a_{\pi\pi}(L)$
& $m_{\pi}a_{\pi\pi}$
& $\xi$
\\
\hline
0.015&0.0265(38)
	&0.00016(16)&0.0780(14)& $-$0.204(23)  &$-$0.200(23)& 0.0545(18)\\
0.025&0.0168(20)
	&$-$0.00003(15)  &0.03056(51)&$-$0.220(22) &$-$0.221(22)&0.0782(15)\\
0.035&  0.0127(16)
	&$-$0.00016(16)  & 0.01415(25) &$-$0.238(25) &$-$0.242(25)&0.1000(21)\\
0.050& 0.0140(21)
	&$-$0.00027(15)  & 0.005535(97) & $-$0.361(45) &$-$0.370(46)&0.1208(22)\\
0.070&  0.0125(19)
	& $-$0.00036(14) & 0.002015(38) & $-$0.447(56)&$-$0.461(57)&0.1507(21)\\
0.100&  0.00836(49)
	 &$-$0.00087(27) & 0.000532(15) & $-$0.453(23)&$-$0.497(28)&0.1828(35)\\
\hline
\end{tabular}
	\label{tab:fse_2}
\end{center}
\end{table}
%%%

\subsection{Chiral Extrapolation}
In this subsection we extrapolate the scattering length to the physical pion mass
from those at the lattice data points obtained in the previous subsection.
The overlap formalism has exact chiral symmetry on the lattice
and the results obtained with this formalism
are expected to be consistent with ChPT even at finite lattice spacings.
Following ref.~\cite{Beane:2005rj}, the scattering lengths for different values of the chiral expansion parameter, $\xi$, are fitted by the expression of the scattering length at the NNLO of ChPT, which is cast in the form
\begin{align}
m_{\pi} a_{\pi\pi}
&= -\pi\xi
\left\{1 +\xi \left(
  \frac{3}{2}\ln\xi  + l_{\pi\pi}^{(1)} \right)
+ \xi^2\left( -\frac{31}{6}(\ln\xi)^2
  +l_{\pi\pi}^{(2)} \ln\xi+ l_{\pi\pi}^{(3)} \right)
\right\} +{\cal O}(\xi^4) , \label{a2l}
\end{align}
as is given in Appendix~\ref{app:ChPT}. 
Here $l_{\pi\pi}^{(i)}$s
are combinations of the low-energy constants in ChPT at 
a quark-mass independent scale.
$l_{\pi\pi}^{(i)}$s are taken as fitting parameters below. 
 We use the scattering lengths at the lightest four to six quark masses for the fit.
 At the NNLO of ChPT, 
$l_{\pi\pi}^{(1)}$ and $l_{\pi\pi}^{(3)}$ are taken as the fitting parameters in eq.~(\ref{a2l}), 
while $l_{\pi\pi}^{(2)}$ is kept to be zero as done in \cite{Noaki:2008iy}.
This is because practically $l_{\pi\pi}^{(2)}$ is hard to be determined.
Instead, we checked how much the fitting parameters change by including $l_{\pi\pi}^{(2)}$
as its phenomenological value and incorporated the changes as systematic errors.
 For comparison, the data are also fitted by the NLO
formula of ChPT with one parameter $l_{\pi\pi}^{(1)}$ by
 truncating ${\cal O}(\xi^3)$ terms in eq.~(\ref{a2l}).

Figure~6 shows the results of the fit of the $m_\pi a_{\pi\pi}$.
The fits at the NLO (left hand side) and the NNLO (right hand side) of ChPT for the six
quark masses are shown.
The leading order (LO) result of ChPT is also shown there.
At a glance we cannot conclude which of the LO, NLO and NNLO of ChPT fits
the calculated $m_\pi a_{\pi\pi}$ best.
Table~IV summarizes the low-energy constants, $l_{\pi\pi}^{(1)}$ and $l_{\pi\pi}^{(3)}$, 
and $m_\pi a_{\pi\pi}$ at the physical pion and the values of $\chi^2$ obtained by the fit of ChPT. 
The values of $\chi^2$ are similar for the LO, NLO and NNLO fits when the results of 
lowest four or five quark masses are used.
When the results of six quark masses are used, however,  the NLO and NNLO fits have similar $\chi^2$ values but the LO fit has larger $\chi^2$ than the NLO and NNLO fits.
This implies that the effect of the NLO and NNLO of ChPT appears only at the largest quark 
mass, $m = 0.100$.

\begin{table}
  \begin{center}
    \caption{Summary of the results of the chiral extrapolation.}
    \begin{tabular}{|c | c |c c c|c|}
	\hline
	order of ChPT & data pts&$l_{\pi\pi}^{(1)}$
	&$l_{\pi\pi}^{(3)}$&$\chi^2 /{\rm d.o.f}$ 
	& $m_{\pi}a_{\pi\pi}$ ({Physical}) \\
	\hline
 	LO & 6   & - & - & 3.1  & $-$0.04539 \\\
 	    & 5   & - & - & 2.2 & $-$0.04539 \\
 	    & 4   & - & - & 2.8  & $-$0.04539 \\
	\hline
 	NLO & 6   & 1.97(23) & - & 1.7  & $-$0.04251(15) \\\
 	    & 5   & 2.39(44) & - & 1.8 & $-$0.04279(29) \\
 	    & 4   & 2.18(60) & - & 2.5  & $-$0.04256(40) \\
	\hline
 	NNLO & 6   & 5.8(1.2) & $-$6.8(6.9) & 1.7 & $-$0.04410(69) \\
 	     & 5   & 5.0(1.9) & 1(16) & 2.2 &  $-$0.0437(11)\\
 	     & 4   & 7.7(4.0) & $-$27(38) & 3.2& $-$0.0451(23) \\
	\hline 
	\end{tabular}
\label{tab:ChPT21}
	\end{center}
\end{table}

In Figure~7 we show $l_{\pi\pi}^{(1)}$ and  $m_\pi a_{\pi\pi}$ at the physical pion mass, which are obtained by the fit at the NNLO and the NLO of ChPT.
In the figure phenomenological values and the results of lattice simulations by other groups are also shown for comparison.
The NNLO fit has larger errors in $l_{\pi\pi}^{(1)}$ and $m_\pi a_{\pi\pi}$ than the NLO fit.
This is because the former has more fitting parameters than the latter.
From Figure~{\ref{fig:lec_and_am}} one sees that the central values of $l_{\pi\pi}^{(1)}$ obtained by the NNLO fit are consistent with the phenomenological value
given in \cite{Colangelo:2001df}, while, those obtained by the NLO fit seem to slightly deviate from the phenomenological values.
These results including a change in convexity of the fitting curve
suggest a significance of the NNLO contributions
especially for the heavier quark masses.
Still, the values of the parameter, $l_{\pi\pi}^{(3)}$, which are determined
using 4, 5 and 6 quark masses, are very much different.
More precise determination of $l_{\pi\pi}^{(3)}$ demands more data points at
lighter quark masses and careful analysis for the
chiral extrapolation of the scattering length.
It is noted, however, that the $m_\pi a_{\pi\pi}$ can be extrapolated to the physical pion mass
with relatively small errors.
The extrapolated scattering length is compared with the newly reported
experimental value extracted from the kaon decay measurement
\cite{PoSNA48, Batley:2007zz}.
It is remarkable that the extrapolated value at the NNLO of ChPT agrees with the result of
ref.~{\cite{PoSNA48}}.
We should also mention that our results are consistent with the results of other groups within errors.

\begin{figure}
\begin{center}
\begin{tabular}{c c}
   \includegraphics[width=8cm,clip=yes]{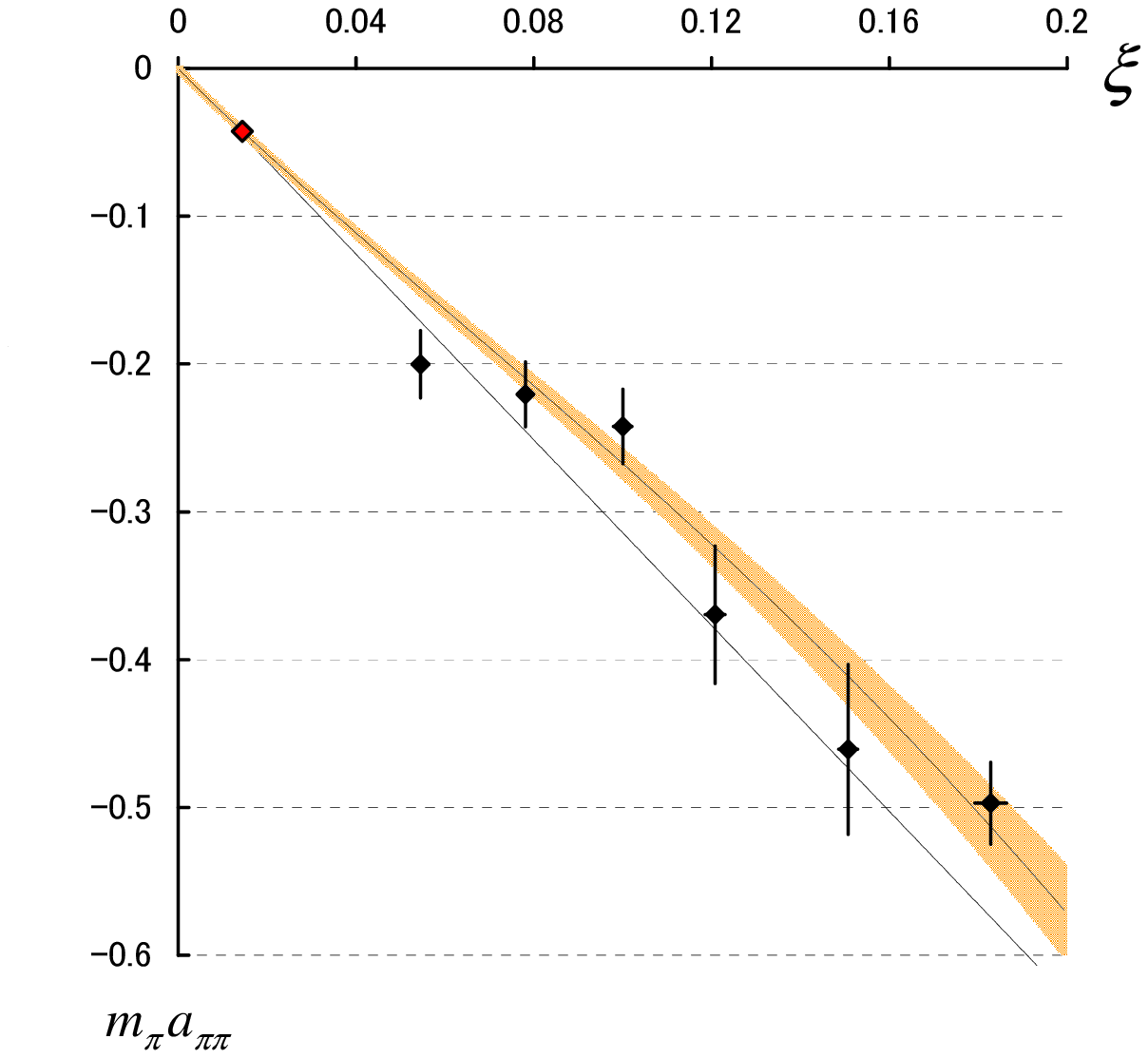} &
   \includegraphics[width=8cm,clip=yes]{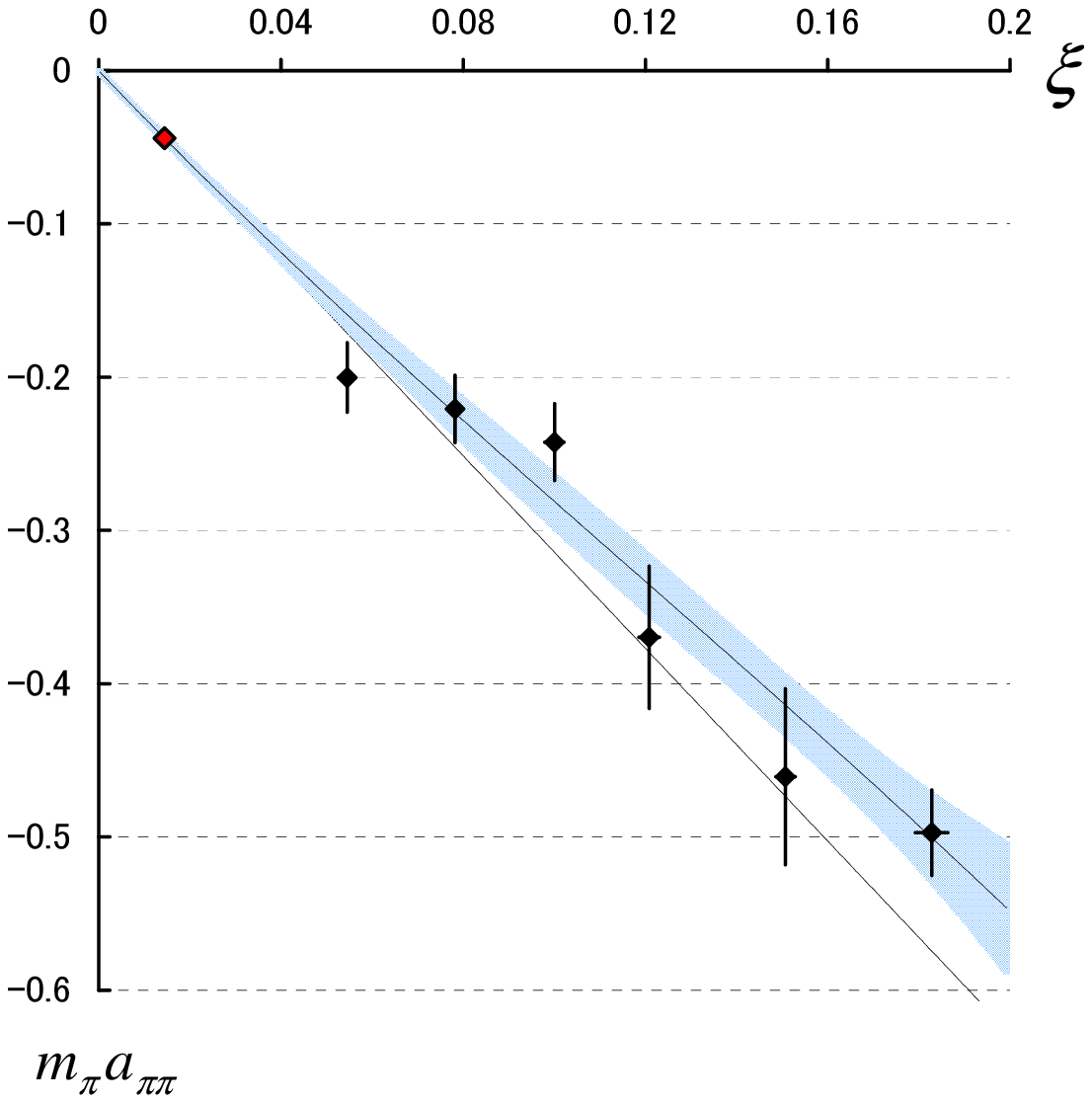} \\
\end{tabular}
    \caption{
ChPT fit at NLO (left figure) and NNLO (right figure) level of
the scattering length for the six quark masses.
The x-axis is the expansion parameter $\xi$, and the y-axis is the
scattering amplitude $m_{\pi}a_{\pi\pi}$.
The straight line from the origin to the lower right is
the leading order (LO) result of ChPT.
In the left (right) figure the curved line shows the NLO (NNLO) fit of ChPT
and the shaded area shows the $1\sigma$ band of the fit.
The upper left point is $m_{\pi}a_{\pi\pi}$ extrapolated to the physical pion mass with using 6 quark masses.
    }
\end{center}
\end{figure}
\begin{figure}
\begin{center}
\begin{tabular}{cc}
\includegraphics[width=8.0cm,clip=yes]{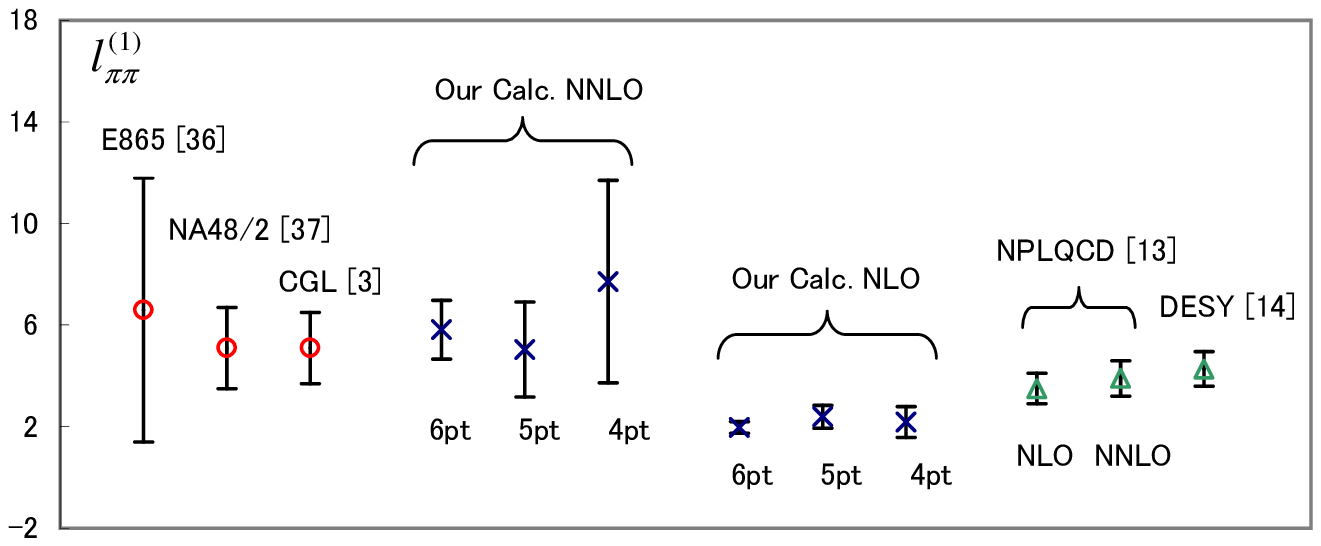}&
\includegraphics[width=8.0cm,clip=yes]{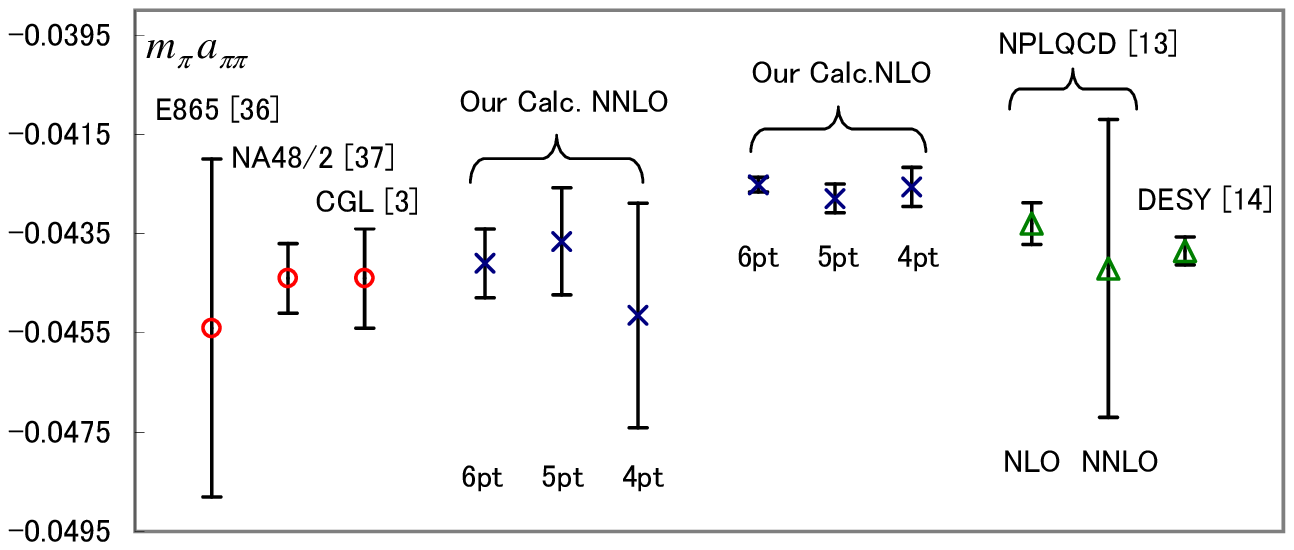}\\
\end{tabular}
    \caption{
  $l_{\pi\pi}^{(1)}$ (left figure ) 
and $m_{\pi}a_{\pi\pi}$ at the physical scale (right figure).
Crosses are our results obtained by the ChPT fit.
The right (left) three points are the results of the NLO (NNLO) fit.
4pt, 5pt, and 6pt below the points mean that 4, 5, and 6 quark masses
are used for the fitting, respectively.
The open circles are phenomenological values obtained by E865 [36], NA48/2 [37] and CGL [3] collaborations and the open triangles are the results of lattice simulations by NPLQCD group [13] and DESY group [14].
}
    \label{fig:lec_and_am}
\end{center}
\end{figure}
\newpage
\section{Summary}

We have calculated the  $I=2$ $\pi\pi$ scattering length using the gauge configurations generated by the JLQCD collaboration with the two-flavor dynamical overlap fermion.
The overlap fermion action has exact chiral symmetry on the finite lattice, 
which enabled us to compare the calculated results with the chiral perturbation theory in the continuum.

We have adopted the technique called low mode averaging in order to gain statistics for the correlation function and have taken into account the wrap-around term in the parametrization of the correlation function in order to identify the contribution of the ground state in the correlation function.

We have also investigated finite volume effects on the physical quantities,
which appear due to pion-loop corrections and also due to fixed global topology.
We have taken into account corrections of these effects on the $\pi\pi$  scattering length.
In order to estimate the correction due to fixed global topology we have used the topological susceptibility obtained by JLQCD for the same gauge configurations.
From our calculation, it is found that the corrections are small
and the lattice volume used for the present calculation is large enough to suppress finite volume effects.

The calculated results are fitted by the NNLO ChPT and also the NLO ChPT for comparison, from 
which the scattering length at the physical pion mass is extrapolated.
It is found that the NNLO ChPT and the NLO ChPT fit the calculated pion mass, the $\pi\pi$ 
scattering length and the low-energy constant of ChPT equally well.
However, the obtained scattering length and the low-energy constant of ChPT at the NNLO ChPT agree slightly better with the experimental ones than those obtained at the NLO ChPT.

The scattering length extrapolated to the physical pion mass assuming the NNLO ChPT is
$m_{\pi} a_{\pi\pi} = -0.04410(69)(18)$, in which the numbers in the first and second parentheses represent the statistical error and the systematic error from using the truncated formula.
The extrapolated scattering length is in good agreement with the experimental result.
A combination of the low-energy constants of ChPT obtained by fitting the calculated results is also consistent with the phenomenological value.

\begin{acknowledgments}
We would like to thank the members of the JLQCD collaboration for allowing us 
to use their gauge configurations for this work. We express our gratitude 
to Jun Noaki in particular for helpful discussions. Numerical simulations are 
performed on IBM System Blue Gene Solution at High Energy Accelerator Research
Organization (KEK) under a support of its Large Scale Simulation Program
(No. 07-13). 
The work of S.H.~and O.M.~was partly supported by Grants-in-Aid for Scientific Research by the Ministry of Education, Culture, Sports, Science and Technology (MEXT) of Japan (Nos. 21674002 and 20540288).
%The work of O.M.~was partly supported by Grants-in-Aid for Scientific Research by the Ministry of Education, Culture, Sports, Science and %Technology (MEXT) of Japan (No. 20540288 and No. 21674002).
\end{acknowledgments}

\appendix
\section{Low Mode Average}
\label{app:low_mode_average}
We decompose the quark propagator into the low mode (``L'') and the high mode (``H'') of the overlap-Dirac operator for each gauge configuration.
Then, for example, the meson correlator is decomposed into four terms according to the two modes of the quark propagator as 
\begin{equation}
  C_{\pi}(t;t_{0}) = 
C_{\rm HH}(t;t_{0}) + C_{\rm HL}(t;t_{0}) + 
C_{\rm LH}(t;t_{0}) + C_{\rm LL}(t;t_{0}).
\end{equation}
Among these four terms, $C_{\rm LL}(t;t_{0})$ is the most important for physical quantities at low energies.
The idea of LMA is to average only $C_{\rm LL}(t;t_{0})$ over the time of the source $t$
\begin{equation}
  \label{eq:lma}
C^{'}_{\pi}(t;t_{0}) = C_{\pi}(t;t_{0}) - C_{\rm LL}(t;t_{0}) + \frac{1}{T}
\sum^{T-1}_{t_{1} = 0} C_{\rm LL}(t - t_{0} + t_{1};t_{1}).
\end{equation}
Eigenvalues and eigenvectors of the low modes of the overlap Dirac operator are stored and are used in order to calculate the average.
We use the same technique for the two-pion state.
\\
\section{finite volume effects from fixed global topology}
\label{app:global_topology}
The $n$-point function with the topological charge $Q$ fixed, is expressed as
\begin{align}
G^{Q}_{n} &= G_{n}(\theta_{s}) + \frac{1}{2\chi_t V}\frac{\rd^2 G_{n}(\theta)}{\rd \theta^2}\Big |_{\theta=\theta_s}
 + \cdots,
\end{align}
where $G_{n}(\theta)$ is the $n$-point function in the $\theta$ vacuum,
\begin{align}
G_{n}(\theta) &= A(\theta)e^{-m_{n}(\theta)t},
\end{align}
where $\chi_t$ is the topological
susceptibility, $V$ is the volume of the lattice and $\theta_s = \frac{iQ}{\chi_t V}$ is the saddle point.
\\

Assuming that  
\begin{align}
m_{n}(\theta) &= m_{n}(0) +\frac{1}{2!}\frac{\rd^2 m_{n}(\theta)}
{\rd \theta^2}\Big |_{\theta=0}\cdot\theta^2 +\cdots,
\end{align}
we can approximate $G^{Q}_{n}$ as
\begin{align}
G^{Q}_{n} &\sim A^{Q}e^{-m^{Q}_{n}t},
\end{align}
where 
\begin{align}
m^{Q}_{n} &= m_{n}(0) + 
\frac{1}{2 V \chi_{t}}
\left( 
1 - \frac{Q^2}{V \chi_{t}}
\right)
\frac{\rd^2 m_{n}(\theta)}{\rd \theta^{2}}
\Big|_{\theta = 0} + {\cal O}(V^{-2}) +\cdots.
\end{align}
Therefore, on the one hand $E_{\pi\pi} - 2m_{\pi}$ is given as
\begin{align}
E_{\pi\pi}^Q - 2m_{\pi}^Q 
&\equiv  m^{Q}_{4} - 2m^{Q}_{2}\\
\label{eq:fse_q_am}
&=  E_{\pi\pi}(0) - 2m_{\pi}(0) +\frac{1}{2 V \chi_{t}}
	\left( 
	1 - \frac{Q^2}{V \chi_{t}}
	\right)\frac{\rd^2 (E_{\pi\pi}(\theta)-2m_{\pi}(\theta))}{\rd \theta^{2}}
	\Big|_{\theta = 0} +  {\cal O}(V^{-2}),
\end{align}
with the fixed topological charge, $Q$.
On the other hand, $E_{\pi\pi} - 2m_{\pi}$ is given from L\"uscher's formula, eq.~(\ref{eq:luscher}),
and the ChPT, eq.~(\ref{aII}), as
\begin{align}
E_{\pi\pi} - 2m_{\pi}
	&= - \frac{4\pi a_{\pi\pi}}
		{m_{\pi}L^3}\left(1 + {\cal O}(L^{-1})\right) \\
	&= \frac{1}{4 f_{\pi}^2 L^{3}}
	\left\{
		1 + \frac{m_{\pi}^2}{16\pi^2 f_{\pi}^2}
		\left(-\frac{3}{2}\tilde{L}  + l_{a} \right)
	\right\} + \cdots. \\
\label{eq:ChPT_delE}
	&=  \frac{1}{4 f^2 L^{3}}
	\left\{
	1 + \frac{m_{\pi}^2}{16\pi^2 f^2} 
	\left(\frac{7}{2}\ln\frac{m_{\pi}^2}{\mu^2}  + l'_{a}\right)
	\right\}+ \cdots,
\end{align}
where
\begin{align}
l'_{a} &= 
 -\frac{4}{3}\bar{l}_{1}
 -\frac{8}{3}\bar{l}_{2}
+\frac{1}{2}\bar{l}_{3}
 -\frac{1}{2}.
\end{align}
In the last line, the following formula for the pion decay constant $f_{\pi}$
is used.
\begin{align}
\label{eq:ChPT_f}
f_{\pi} &=   f
\left\{1 -  \frac{m_{\pi}^2}{16\pi^2 f_{\pi}^2} \left(
  -\tilde{L}  - \bar{l}_{4} \right)\right\}
  + \cdots,
\end{align}
where $f$ is the pion decay constant in the chiral limit.
Then, we assume that the $\theta$-dependence of $E_{\pi\pi} - 2m_{\pi}$ appears only through
that of the pion mass, which is taken into account by changing $m_{\pi}$ into the following form of $m_{\pi}(\theta)$
\begin{align}
m_{\pi} \rightarrow m_{\pi}(\theta) &= 2B_{0}m_{q}\cos(\theta/N_{f}),
\end{align}
where  $m_{q}$ is the quark mass and $N_{f}$ is the
number of the flavor.
By taking the second derivative of eq.~(\ref{eq:ChPT_delE}) by $\theta$, substituting it
into eq.~(\ref{eq:fse_q_am}) and setting $Q=0$, 
we obtain the following expression for the finite volume correction due to the fixed global topology:
\begin{align} 
\delta(E_{\pi\pi} - 2m_{\pi}) &\equiv
(E_{\pi\pi}^Q - 2m_{\pi}^Q)
-(E_{\pi\pi}(0) - 2m_{\pi}(0))\\
&=
\frac{1}{4f^2 L^3}
\left[
\frac{1}{8V \chi_{t}} 
 \frac{m_{\pi}^2(0)}{16 \pi^2 f^2 }
\left\{\frac{7}{2}
\left(
 \ln\frac{m_{\pi}^2 (0)}{\mu^2} +1
 \right)
 + l'_{a}(\mu)
\right\}
  + {\cal O}(V^{-2})
\right].
\end{align}
\section{Scattering length in ChPT at NNLO}
\label{app:ChPT}
The $\pi\pi$ scattering length is given at the NNLO of ChPT in ref.~\cite{NPB508}.
The S-wave scattering length in $I=2$ channel is expressed as
\begin{align}
m_{\pi}a_{\pi\pi} &= -\pi\xi
	\left\{
	1 - \xi[2 + \bar{b}_{1} + 16\bar{b}_{4}] + \xi^2
	\left[	
	\frac{262}{9} - \frac{22\pi^{2}}{9} + 4\bar{b}_{1}
	+ 64\bar{b}_{4}
	\right]
	\right\}+{\cal O}(\xi^4), \label{aChPT}
\end{align}
where $\bar{b}_i$'s are coefficients (multiplied by $16\pi^2$)
introduced in ref.~\cite{PLB,NPB508,Colangelo:2001df} 
to parametrize the pion scattering amplitude.
$\bar{b}_i$'s are
\begin{align}
\bar{b}_{1} + 16\bar{b}_{4}
 = & \ \frac{3}{2}\tilde{L} +
\frac{4}{3}\tilde{l}_1+\frac{8}{3}\tilde{l}_2-\frac{1}{2}\tilde{l}_3
-2\tilde{l}_4-\frac{3}{2}  + \frac{31}{6}\xi\tilde{L}^2
+\xi\tilde{L}
\left(\frac{4}{3}\tilde{l}_1+8\tilde{l}_2-\tilde{l}_3+2\tilde{l}_4-\frac{47}{12}
\right) \nonumber \\ &
+\xi\left(
\frac{16}{3}\tilde{l}_1\tilde{l}_4+\frac{32}{3}\tilde{l}_2\tilde{l}_4
-3\tilde{l}_3\tilde{l}_4-5\tilde{l}_4^{\ 2}-\frac{1}{2}\tilde{l}_3^{\ 2}
+4\tilde{l}_1+\frac{16}{3}\tilde{l}_2
-\frac{15}{4}\tilde{l}_3-6\tilde{l}_4+\frac{1861}{144}
+\tilde{r}_1+16\tilde{r}_4
\right),  \label{LECb}
\end{align}
with the low-energy constants $\tilde{l}_i$, $\tilde{r}_i$ in ChPT 
of chiral order 4 and 6 renormalized at scale $\mu$ and
\[
 \xi= \frac{m_\pi^2}{16\pi^2 f_\pi^2} , \ \ \tilde{L}=-\ln\frac{m_\pi^2}{\mu^2}.
\]

Substituting eq.~(\ref{LECb}) into eq.~(\ref{aChPT}) and
rearranging the result in the order of $\xi$ lead to the following
expression:
\begin{align}
m_{\pi} a_{\pi\pi} 
&= -\pi\xi
\left\{1 +\xi \left(
  -\frac{3}{2}\tilde{L}  + l_{a} \right)
+ \xi^2\left( -\frac{31}{6}\tilde{L}^2
  + l_{b} \tilde{L}+ l_{c} \right)
\right\}+{\cal O}(\xi^4), \label{aII}
\end{align}
with
\begin{align}
l_{a} &= 
 -\frac{4}{3}\tilde{l}_{1} -\frac{8}{3}\tilde{l}_{2} +\frac{1}{2}\tilde{l}_{3} +
 2\tilde{l}_{4} -\frac{1}{2}, \nonumber\\
l_{b} &= 
{-}\frac{4}{3}\tilde{l}_{1} {-} 
	8\tilde{l}_{2} {+}
	\tilde{l}_{3}  {-}
	2\tilde{l}_{4} + \frac{119}{12}, \nonumber\\
l_{c} &={\frac{1}{2}\tilde{l}_3^{\ 2}-}
		\left(
			\frac{16}{3}\tilde{l}_{1} + \frac{32}{3}\tilde{l}_{2}
			-3\tilde{l}_{3} -5\tilde{l}_{4}
		\right)	\tilde{l}_{4} +
	\frac{4}{3}\tilde{l}_{1}  +
	\frac{16}{3}\tilde{l}_{2} +
	\frac{7}{4}\tilde{l}_{3} -
	2\tilde{l}_{4} +{\frac{163}{16}}
	- \frac{22}{9}\pi^2 {-} \tilde{r}_1-16 \tilde{r}_4.
\end{align}
The scale dependence of $\tilde r_i$ is fixed so that $\bar{b}_i$ has no scale dependence,
$\mu d\bar{b}_i/d\mu=0$.
Therefore, the right hand side of eq.~(\ref{aChPT}), 
is scale independent, though truncated at the third order of $\xi$.
Then, the right hand side of eq.~(\ref{aII}) as a whole is also scale independent.
Thus, one can choose $\mu$ arbitrarily.
When we fit the scattering length obtained from lattice simulations as a function of $\xi$,
we want to make the fitting parameters quark-mass independent.
Therefore, we choose $\mu=4\pi f$ with the pion decay constant in the chiral limit $f$
At this scale, it should be noted that
\begin{align}
  \tilde{L}(\mu=4\pi f) & = -\ln \xi -\ln{\frac{f_\pi^2}{f^2}} \nonumber \\
& = -\ln\xi -2 \xi \bar{l_4} +{\cal O}(\xi^2) \nonumber \\
& = -\ln\xi -2 \xi \tilde{l}_4(\mu=4\pi f)+2\xi\ln\xi +{\cal O}(\xi^2),
\label{ChLn}
\end{align}
in which we have used the chiral expansion of the pion decay constant 
$f_\pi=f\{1+\xi \bar{l}_4 +{\cal O}(\xi^2)\}$
with $\bar{l}_4$ being the low-energy constant at the pion-mass scale
\cite{NPB508,Colangelo:2001df}.

Substituting eq.~(\ref{ChLn}) into eq.~(\ref{aII}), setting the scale $\mu=4\pi f$
and sorting the result in order of $\xi$,
we obtain the following expression for the pion scattering length
$^{\rm \footnotemark[2]}$
\footnotetext[2]{Note that the sign of the coefficient of $(\ln\xi)^2$ is different from
that of eq.~(17) in [4].}
\begin{align}
m_{\pi} a_{\pi\pi}
&= -\pi\xi
\left\{1 +\xi \left(
  \frac{3}{2}\ln\xi  + l_{a} \right)
+ \xi^2\left( -\frac{31}{6}(\ln\xi)^2
  -(l_{b}+3) \ln\xi+ (l_{c}+3\tilde{l}_4) \right)
\right\} +{\cal O}(\xi^4). \label{aIII}
\end{align}
Note that all the low-energy constants are at the scale of
$\mu=4\pi f$ and are independent of the quark mass, so that we can take them
as fitting parameters in the chiral extrapolation of the scattering length.

\end{document}